\documentclass[10pt]{wlscirep}

\newlength{\twin}
\newlength{\figwidth}
\usepackage[utf8]{inputenc}
\usepackage{textcomp,marvosym}
\usepackage{amsmath,amssymb}
\hypersetup{urlcolor=black}
\usepackage{graphicx}
\usepackage{float}
\usepackage[caption = false]{subfig}
\usepackage{sidecap}

\usepackage{booktabs}
\usepackage{tabularx}
\usepackage{a4wide}
\newcommand{\expect}[1]{
	\langle #1 \rangle
}
\newcommand{\pp}[1]{\left(#1\right)}
\usepackage{xcolor}
\definecolor{red}{rgb}{0.8, 0.0, 0.0}

\title{Evidence of Complex Contagion of Information in Social Media: An Experiment Using Twitter Bots}

\author[1]{Bjarke Mønsted}
\author[1]{Piotr Sapieżyński}
\author[2]{Emilio Ferrara}
\author[1,*]{Sune Lehmann}
\affil[1]{Technical University of Denmark, Department of Applied Mathematics and Computer Science, 2800 Lyngby, Denmark}
\affil[2]{Information Sciences Institute, University of Southern California, Marina del Rey, CA 90292, USA}

\affil[*]{\href{mailto:sljo@dtu.dk}{sljo@dtu.dk}}

\begin{abstract}
It has recently become possible to study the dynamics of information diffusion in techno-social systems at scale, due to the emergence of online platforms, such as Twitter, with millions of users. 
One question that systematically recurs is whether information spreads according to simple or complex dynamics: does each exposure to a piece of information have an independent probability of a user adopting it (simple contagion), or does this probability depend instead on the number of sources of exposure, increasing above some threshold (complex contagion)?
Most studies to date are observational and, therefore, unable to disentangle the effects of confounding factors such as social reinforcement, homophily, limited attention, or network community structure. 
Here we describe a novel controlled experiment that we performed on Twitter using `social bots' deployed to carry out coordinated attempts at spreading information. 
We propose two Bayesian statistical models describing simple and complex contagion dynamics, and test the competing hypotheses. 
We provide experimental evidence that the complex contagion model describes the observed information diffusion behavior more accurately than simple contagion.
Future applications of our results include more effective defenses against malicious propaganda campaigns on social media, improved marketing and advertisement strategies, and design of effective network intervention techniques.
\end{abstract}
\begin{document}

\flushbottom
\maketitle
%
%
\thispagestyle{empty}

\section*{Introduction}
The diffusion of information and ideas in complex social systems has fascinated the research community for decades~\cite{castellano2009statistical}.
The first proposal to use epidemiological models for the analysis of the spreading of ideas was put forth more than fifty years ago~\cite{goffman1964generalization}. 
Such models, where each exposure results in the same adoption probability, are referred to as \textit{simple contagion} models.

It was subsequently suggested, however, that more complex effects might come into play when considering the spread of ideas rather than diseases.
For example, some people tend to stop sharing information they consider ``old news'', while others refuse to engage in discussions or sharing certain opinions they do not agree with~\cite{granovetter1978threshold, daley1964epidemics, daley1965stochastic}.
Such models, in which adoption probabilities instead depend strongly on the number of adopters in a person's social vicinity in a way where exposure attempts cannot be viewed as independent, are referred to as \textit{complex contagion}~\cite{centola2007complex} models. 
Concretely, we use a threshold complex contagion model, in which the adoption probability is assumed to increase slowly for low number of unique exposure sources, then increase relatively quickly when the number of sources approaches some threshold level (see `Models' for full details).

The role of contagion in the spreading of information and behaviors in (techno-)social networks is now widely studied in computational social science~\cite{centola2010spread, onnela2010spontaneous, romero2011differences, bond201261, ugander2012structural, weng2013virality, muchnik2013social, karsai2014complex, kramer2014experimental, hodas2014simple, ferrara2015measuring, goel2015structural, lerman2016information}, with applications ranging from public health~\cite{campbell2013complex} to national security~\cite{subrahmanian2016darpa}. 
The vast majority of these studies are, however, either observational, and therefore prone to biases introduced by confounding factors (network effects, cognitive limits, etc.), or entail controlled experiments conducted only on small populations of a few dozens individuals~\cite{centola2007complex, centola2010spread}. 
To date, these limitations have prevented the research community from drawing a conclusive answer as to the role of simple and complex information contagion dynamics at scale.

In this paper we shed new light on the nature of information diffusion using a large-scale experiment on Twitter, in which we study the spreading of hashtags within a controlled environment. 
Creating a controlled environment for experiments within online platforms is especially challenging for researchers that do not have access to the system's design itself, as traditional techniques such as A/B testing cannot be employed. 
Even for service providers like Facebook, ethical concerns emerged when random control trials were carried out without review board approval~\cite{kramer2014experimental}.

For this experiment, we leveraged algorithm-driven Twitter accounts (social bots)~\cite{ferrara2016rise}. 
We had previously shown that a coordinated network of Twitter bots can be effective in influencing trending topics on Twitter~\cite{Sapiezynski_2014_conference}.
This study is a follow-up experiment designed to quantitatively investigate how users react to information stimuli presented by single or multiple sources.
In particular, for this experiment, teams of students from the Technical University of Denmark (DTU) worked together to create a network of Twitter bots (a botnet) designed to attract a large number of human followers. 
We programmed the bots to spread Twitter hashtags (see Table~\ref{tab:hashtags}) in a synchronized manner among a set of real Twitter users from a selected geographical area.
A large number of users in our target dataset followed one or multiple bots (See Figure~\ref{fig:followers_vs_time}B), which allowed us to study the effect of multiple exposures from distinct sources on information contagion. 

The decision to use Twitter bots to perform coordinated interventions has several advantages: first, we are able to ensure that the hashtags we introduce are new to Twitter, and therefore that they are seen by the target users for the first time when we perform experiments. Second, it enables the bots to work together to expose users to each intervention multiple times.
Finally, the Twitter botnet mitigates the confounding effects of homophily~\cite{mcpherson2001birds, aral2009distinguishing, shalizi2011homophily}.
For example, when conducting a purely observational study, it is a fundamental problem to distinguish whether a user is more likely to adopt information shared by many of their friends because they are influenced by their friends sharing the content, or simply because friends tend to be similar, so anything tweeted shared by the user's friends is more likely to be of interest to the user.

In the remainder of the paper we will discuss the experimental framework design in detail, then present two statistical models for simple and complex contagion, developed in order to evaluate the two competing hypotheses, and finally show the results of the experimental evaluation.

\section*{Results}
\paragraph{Deploying the botnet.}
Creating a botnet with a large number of followers with a network structure suited for testing our hypotheses presented several challenges which are described below. 

We began by ensuring that the bots would appear to be human-like if subjected to a cursory inspection. 
We achieved this goal by having the bots generate content using simple natural language processing rules as well as `recycling' popular content from other Twitter users. 
We also had the bots tweet at irregular intervals, but with frequencies set according to a circadian pattern. 
Finally, we used some Twitter users' tendency to reciprocate friendships to ensure that the bots were followed by a large number of accounts while themselves following only a few; a following/follower ratio much smaller than one is unusual in typical twitter bots. 
The full botnet consisted of 39 algorithmically driven Twitter accounts.
See `Materials and Methods' for full details on botnet-creation. 

\begin{figure}
	\centering
	\includegraphics[width=0.75\linewidth]{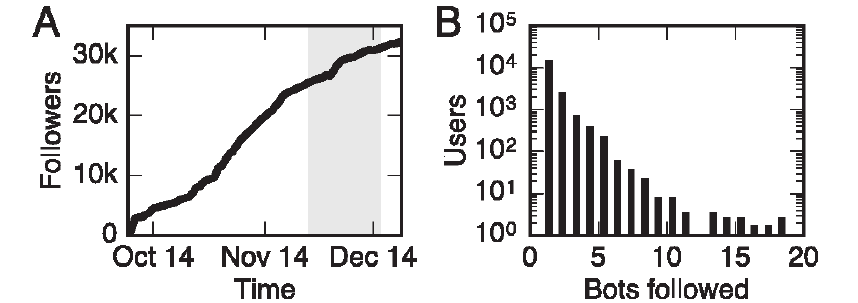}
	\caption{Illustration of the status of our botnet at the time of the interventions. The bots had accumulated a large number ($\sim\!25\,000$) of followers (A) at the time of the interventions (shaded region), and many of the target users followed several distinct bots (B).}
	\label{fig:followers_vs_time}
\end{figure}

Once we had established the botnet, we focused on establishing a network structure that would allow for investigating the mechanism driving contagion processes. 
Our strategy was simple: Whenever a user followed one of our bots, the ID of this user was automatically communicated to the remaining bots, which then also attempted to get that user to follow them. 
This strategy resulted in a botnet followed by a large number of human users (around 25\,000 total followers a the time of interventions), in which a large users followed multiple bots, allowing us to test the effects of multiple exposures to information. 
Figure~\ref{fig:followers_vs_time}A shows the total number of followers as a function of time, while Figure~\ref{fig:followers_vs_time}B displays the distribution of users following $n$ bots. 
Having obtained a botnet with a large number of followers and a desirable network structure, the bots performed a series of coordinated interventions, described in the following.

The general intervention strategy implemented by the bots follows:
\begin{enumerate}
	\item Each bot tweets 2 original tweets about a given \#hashtag;
	\item Each bot retweets the first 4 tweets about that \#hashtag;
	\item Each bot retweets 15 tweets containing that \#hashtag that do not originate from other bots;
	\item Each bot favorites all tweets about the given \#hashtag.
\end{enumerate}

\begin{figure}[t]
	\centering
	\includegraphics[width=\linewidth]{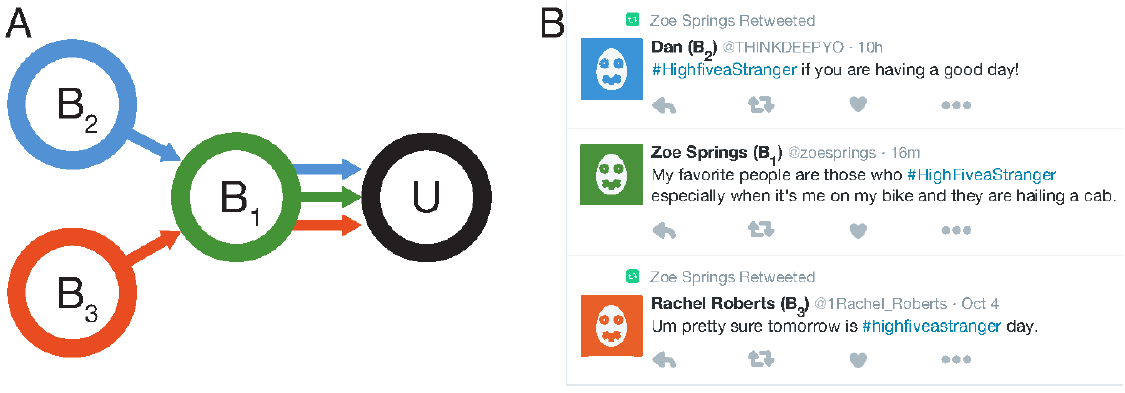}
	\caption{Bots in a botnet can work together to provide users with multiple exposures to an intervention. (A) user $U$ only follows bot $B_1$. Bot ($B_1$) acts as a proxy and exposes the user not only to its own content, but also to content from two other bots ($B_2$ and $B_3$), that the user does not follow. (B) Twitter feed from the perspective of user $U$.}
	\label{fig:proxy_illustration}
\end{figure}

Step 1 of this protocol was based on human-generated tweets; this allowed students to create content designed to increased the likelihood of adoption.
Steps 2--4 were instead automated.
By retweeting each other's content, the bots provided a higher exposure to the target users with respect to what would have been possible if bots could only have targeted their mutual friends, as illustrated in Figure \ref{fig:proxy_illustration}.
An overview of the hashtags that we introduced is shown in Table \ref{tab:interventions}.
The hashtags we introduced support positive behaviors (e.g., encouraging vaccinations or positive human interactions, sharing something good, etc.) and in some cases are contextualized with the time period of the intervention (e.g., fostering stories about Thanksgiving and Black Friday).

\begin{table}[!t]
	\renewcommand{\arraystretch}{1.3}
	\caption{List of interventions}
	\label{tab:interventions}
	\centering
	\begin{tabularx}{\linewidth}{lX}
		\toprule
		Hashtag & Message \\
		\midrule
		\#getyourflushot & Encouraging  Twitter users to vaccinate.\\
		\#highfiveastranger & Encouraging users to engage in positive human interactions.\\
		\#somethinggood & Sharing a recent positive experience.\\
		\#HowManyPushups & Encouraging healthy behaviors and fitness.\\
		\#turkeyface & Photoshopping a celebrity's face onto a turkey.\\
		\#SFThanks & Hashtag for Thanksgiving in San Fransisco.\\
		\#blackfridaystories & Sharing Black Friday shopping stories.\\
		\#BanksySF & Rumor that Banksy, the street artist, was in San Fransisco.\\
		\bottomrule
	\end{tabularx}\label{tab:hashtags}
\end{table}

To track exposures and contagions, each bot automatically recorded when a target user retweeted intervention-related content, and also each exposure that had taken place prior to the retweeting. 
It is important to remark that users cannot be expected to consume the entirety of content generated by those they follow: the probability of seeing a tweet can depend on many factors including the total number of accounts a user follows, the activity level of each of those accounts, and the amount of time that user spends on Twitter.
The two contagion models we created, described in the following, model this uncertainty explicitly.

\subsection*{Models}
In the following, we propose two contagion models, namely \textit{simple contagion} model (SC), in which all exposure attempts are considered to be independent, and \textit{complex contagion} threshold model (CC), and derive quantitative predictions for them. 
Both models take into account the uncertainty regarding the target users observing a given tweet. 
Specifically, we do not have direct access to a user's \textit{actual exposures}, to an intervention, but only to the \textit{attempted exposures}, $N$.
The simple contagion model employs only the total number of attempted exposures, which we denote $k$. The complex contagion model, however, is only concerned with the number of unique sources $\kappa$ from which one or more exposures have succeeded. This is because a central idea in threshold models is the reluctance to partake in activities until a number of individuals in one's social group have already done so~\cite{granovetter1978threshold}. To avoid cluttered notation, we write $k$ and `number of exposures' in descriptions relevant for both models, although these should be replaced with $\kappa$ and `number of unique exposure sources' for the CC case.

Going forward, we separate the two factors that enter into a users tendency to adopt a behavior. 
Firstly, the probability of the user experiencing $k$ exposures, and, secondly, the probability $P(\textrm{RT}|k)$ of the user deciding to retweet content after experiencing $k$ exposures. 
Thus, we model the probability of a user retweeting content from an intervention, given bot activity $A$ as
\begin{equation}
P(\textrm{RT}|A) = \sum_k P(k | A)P(\textrm{RT}|k).\label{eqn:general_model}
\end{equation}
In the following, $A = [a_1, a_2,\ldots]$ denotes a list of the number of times a user has received attempted exposures from each bot (disregarding those with zero attempts). 
For example, $A = [2, 1, 3, 1]$ means that a given user has been subjected to 2 attempted exposure from one bot, 3 from another bot, and 1 from two bots. 
In the case of SC, where only the total number of exposures is of interest, we will use $N = \sum_i a_i$ to denote the total number of attempted exposures.

\paragraph{Simple Contagion.}
We model the number of exposures by assuming that a user sees a given tweet with some independent probability $q$. 
Thus, the number of actual exposures follows a binomial distribution $B(k; N, q)$ given by $q$ and the number of attempted exposures $N$,
\begin{align}
P(k|N,q) = \begin{pmatrix}N \\ k\end{pmatrix} q^k(1-q)^{N-k} = \frac{N!}{k!(N-k)!}q^k(1-q)^{N-k}. \label{eqn:p_k_sc}
\end{align}
In SC, each actual exposure has some probability $\rho$ of `infecting' the user, which is independent of other exposures. Hence the probability for an infection after $k$ exposures is simply
\begin{equation}
P(\textrm{RT}|k)_\text{SC} = 1 - (1 - \rho)^k, \label{eqn:p_rt_given_k_sc}
\end{equation}
which is almost linear in $k$ for small values of $\rho$. 
Inserting this expression into equation~\eqref{eqn:general_model} we get
\begin{equation}
P(\textrm{RT} | N)_\text{SC} = \sum_{k=0}^N \frac{N!}{k!(N-k)!}q^k(1-q)^{N-k} (1 - (1 - \rho)^k), \label{eqn:p_rt_sc}
\end{equation}
which is equivalent to the simpler expression
\begin{equation}
P(\textrm{RT} | N)_\text{SC} = 1 - (1 - \rho q)^N. \label{eqn:sc_model}
\end{equation}
Under \textit{results}, we fit the parameters in equation \eqref{eqn:sc_model} to the data obtained by our experiment. Next, we derive an expression for the retweet probability of the complex contagion model.

\paragraph{Complex Contagion.}
When quantifying the predictions of CC, we face two obstacles: \textit{(i)} redefining the conditional retweet probability $P(\textrm{RT}|A)$ in order to incorporate the threshold effect of CC; and, \textit{(ii)} obtaining an expression of the probability distribution for $\kappa$ given the previous activity $A$. 

Let us first derive the probability distribution for $\kappa$ given previous activity $A$.
The probability $p_i$ of source $i$ resulting in one or more actual exposures is given by a binomial distribution using similar considerations as those leading to equation~\eqref{eqn:sc_model}:
\begin{equation}
p_i = 1 - (1 - q)^{a_i}.
\end{equation}
Hence, the distribution of unique exposures is the result of independent draws from $|A|$ Bernoulli trials with $a_i$ draws from each, with individual success probabilities $p_i$, also known as Poisson's Binomial~\cite{wang1993number}. 
For example, the probability of $\kappa=1$ given a list of attempted exposures $A$ is obtained by summing over the different ways we may achieve success in only a single Bernoulli trials:
\begin{equation}
P(\kappa=1|A) = \sum_{j = 1}^{|A|}\overbrace{\left(1 - (1-q)^{a_{j}}\right)}^{p_{j}}\prod\limits_{i\ne j}\overbrace{\left(1 - q\right)^{a_i}}^{1-p_i}.
\end{equation}
Generalizing this to any $\kappa \leq |A|$, we sum over every unique combination of $\kappa$ successful trials. 
Denoting the set of sets of $\kappa$ integers between $1$ and $|A|$ by $S_\kappa$, we get
\begin{align}
P(\kappa | A) &= \sum_{s\in S_\kappa} \prod\limits_{i\in s}\left(1 - (1-q)^{a_i}\right)\prod\limits_{i \notin s} (1-q)^{a_i}, \label{eqn:p_k_cc}\\
S_\kappa &= \left\{s \subseteq \left\{1, 2, \ldots, |A| \right\}, |s| = \kappa \right\}. \label{eqn:subsets}
\end{align}
We include a note in the SI on how to efficiently compute equation~\eqref{eqn:p_k_cc}, as this expression becomes infeasible to compute using a brute force approach when $|A| > ~25$.
As in equation~\eqref{eqn:general_model}, we sum over positive $\kappa$ to obtain a final expression for the retweet probability given a list $A$ of exposure attempts, by computing the sum $\sum_\kappa P(\kappa|A)P(\textrm{RT}|\kappa)$ over the probabilities given by equation~\eqref{eqn:p_k_cc}.
\begin{align}
P(\textrm{RT}|A) = \sum_{\kappa=1}^{|A|} \sum\limits_{s\in S_\kappa} \prod\limits_{i\in s} \left(1 - (1-q)^{a_i}\right)\prod\limits_{i\notin s}(1-q)^{a_i}P(RT|\kappa). \label{eqn:p_rt_cc}
\end{align}
Now we select a threshold function for $P(RT|\kappa)$. 
We choose a Sigmoid function,
\begin{equation}
P(\textrm{RT}|\kappa)_\text{CC} = \rho_l + \frac{\rho_h - \rho_l}{1 + e^{-w(\kappa-\kappa_0)}},\label{eqn:p_rt_given_k_cc}
\end{equation}
as it employs both a threshold $\kappa_0$, steepness $w$ and the lower and upper limits, $\rho_l$ and $\rho_h$. Sigmoids are commonly used to model soft thresholds, for example as activation functions in neural networks~\cite{basheer2000artificial}, or as fuzzy membership functions~\cite{shi1999implementation}.
Combining \eqref{eqn:p_rt_cc} and \eqref{eqn:p_rt_given_k_cc}, the expression for $P(\textrm{RT}|A)$ becomes
\begin{equation}
P(\textrm{RT}|A)_\text{CC} = \sum_{\kappa=1}^{|A|} \sum\limits_{s\in S_\kappa} \prod\limits_{i\in s} \left(1 - (1-q)^{a_i}\right)\prod\limits_{i\notin s}(1-q)^{a_i}\left(\rho_l + \frac{\rho_h - \rho_l}{1 + e^{-w(\kappa-\kappa_0)}}\right). \label{eqn:cc_model}
\end{equation}
Having derived expressions for the retweet probability for a user given previous exposure activity for both the SC and CC hypotheses, we proceed to fit the models to our experimental data.

\paragraph{Analysis.}
We now use these two contagion models to investigate how the adoption probability $P(\textrm{RT})$ varies as a function of the exposure numbers in our dataset.
By studying how well each model fits the observed data, we can determine which model is the most appropriate description of the contagion processes measured in the experiment. 

\begin{figure}[t!]
	\centering
	\includegraphics[width=0.5\linewidth]{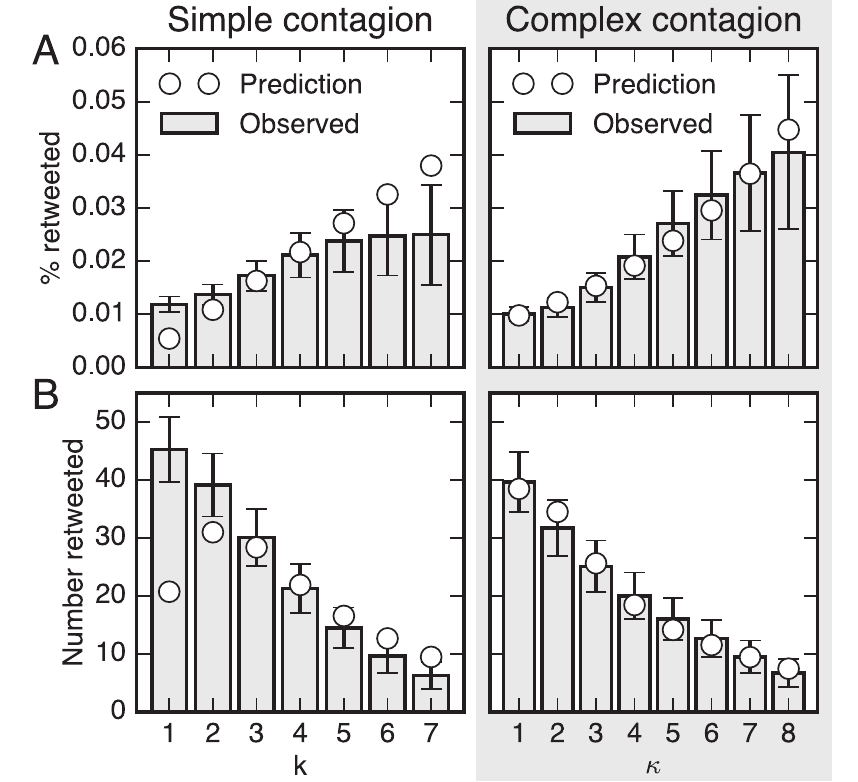}
	\caption{Simple contagion (SC, left) does not adequately describe the contagion dynamics: the best fit underestimates the probability of retweeting after a low number of exposures and overestimates the probability with a large number of exposures. The best fit of complex contagion (CC, right) dynamics correctly estimates the probability of retweeting across the number of sources of exposure. A. Percentage of tweets that were retweeted after $k$ successful exposures (SC) or after exposures from $\kappa$ sources (CC). B. Number of tweets retweeted following $k$ successful exposures (SC), or after exposures from $\kappa$ sources (CC). Best fit of SC model (equation \ref{eqn:p_rt_sc}) and CC model (equation~\ref{eqn:p_rt_cc}) to the data using $q = 0.20$, plotted up to $k=7$ (and $\kappa=8$) to avoid plotting noisy data for large values of $k$ (and $\kappa$).}
	\label{fig:sc_cc_fit}
\end{figure}

An example of the distributions for $q = 0.2$ and the best fits of the SC and CC models are shown in Figure~\ref{fig:sc_cc_fit}.

Figure \ref{fig:sc_cc_fit} suggests that the SC model from~\eqref{eqn:p_rt_sc} is not an adequate fit to the observed data, whereas the CC model from~\eqref{eqn:p_rt_cc} provides an excellent fit. The figure indicates that the CC model, which models contagion as a function of the number of distinct sources provides a better explanation for the user behavior on Twitter.

\begin{figure}[t!]
	\centering
	\includegraphics[width = 0.75\linewidth]{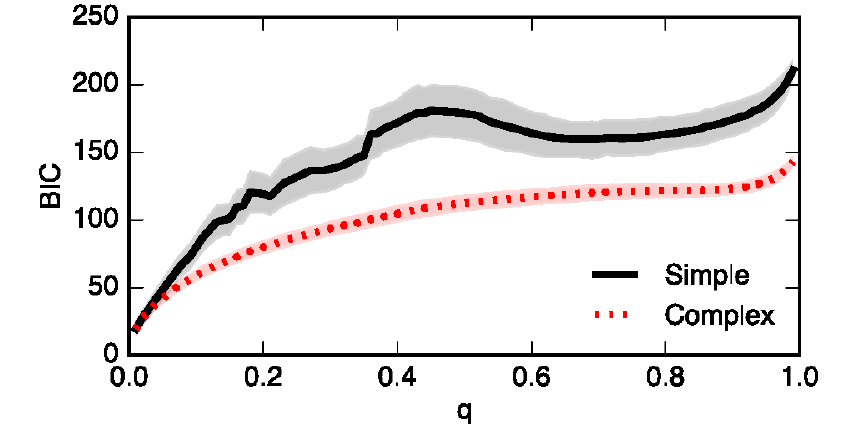}
	\caption{BIC scores for both SC and CC models for a range of values of $q$, the lower the score the better. Across the values of the $q$ parameter, complex contagion model achieves lower BIC scores than simple contagion. The thick lines are the mean values of the simulations, and the shaded regions are the percentiles corresponding to one standard deviation, i.e. they contain $68\%$ of the simulation results.}
	\label{fig:BIC}
\end{figure}

In order to compare the models in a way that takes into account different model complexities, we use the Bayesian Information Criterion (BIC) score~\cite{Schwarz1978} on simulations using the probabilities provided by the two models (see Methods for details).
The results, displayed in Figure~\ref{fig:BIC}, show that the CC model results in better BIC scores for any value of $q$. In general, a difference in BIC scores larger than 10 points is considered a very strong evidence in support of the model with the lower score~\cite{Raftery1995}. Figure~\ref{fig:BIC} shows gaps between the average BIC scores of the two models that are substantially larger than 10 points throughout the entire range of values of $q$, supporting the hypothesis that the CC model is the best explanation for the dynamics of information diffusion on Twitter.

For very small values of $q$ ($q<0.1$) the gap between the BIC scores of the two models is small, and as $q$ grows the gap increases to reach its maximum for values of $q$ around $0.5$. 
The reason for the low BIC scores in the case of very low values of $q$ is that the estimates of exposure numbers from equations \eqref{eqn:p_k_sc} (SC) and \eqref{eqn:p_k_cc} depend on $q$ and yield a very low number of estimated successful exposures for low values of $q$, which causes the error bars on the number of estimated retweets (such as those from Figure~\ref{fig:sc_cc_fit}B) to grow large.

\section*{Discussion}
Diffusion phenomena in social and techno-social systems have attracted much attention due to the importance of understanding dynamics such as disease propagation, adoption of behaviors, emergence of consensus and influence, and information spreading~\cite{castellano2009statistical, centola2007complex, centola2010spread, onnela2010spontaneous}.
In contrast to modeling epidemics, for which clear laws have been mathematically formulated and empirically validated~\cite{goffman1964generalization, daley1964epidemics}, modeling and understanding information diffusion has proved challenging, in part due to the inability to perform controlled experiments at scale and due to the abundance of confounding factors that bias observational studies~\cite{mcpherson2001birds, aral2009distinguishing, shalizi2011homophily}. 
Two competing hypothesis have been debated, namely that information spreads according to simple or complex contagion. 
In this work we test the two hypotheses by creating a controlled experimental framework on Twitter: we deployed 39 coordinated social bots~\cite{ferrara2016rise} that interacted with a selected cohort of participants (our target population), and carried out a variety of interventions, in the form of attempts to spread new positive messages (i.e., memes for social good). The bots recorded the behavior of the target users and all their interactions with the bots and with other users, while tracking the number of exposures to each message over a period of more than one month.
The data we collected allowed us to test two Bayesian models that we derived to capture the diffusion dynamics of simple and complex information contagion. Specifically, in our complex contagion model, we assume that the probability of adoption depends on the number of unique sources of information, rather than the number of exposures.

The statistical evidence clearly shows that the complex contagion model is a better explanation for the observed data than the simple contagion model. 
This imples that exposures from multiple sources impacts the probability of spreading a given piece of information. 
This threshold mechanism differs significantly from, say, the spreading of a virus, where many exposures from a single source are sufficient to increase probability of infection.
A variety of explanations for the complex contagion hypothesis have been proposed in social theory, including social reinforcement and social influence, echo chambers, human cognitive limits, etc.~\cite{castellano2009statistical, granovetter1978threshold, romero2011differences, bond201261, ugander2012structural, muchnik2013social, lerman2016information}. 
While our work identifies the type of mechanism according to which information spreads from person to person, much work is still needed to discriminate which factors drive this phenomenon. 
We expect that future work will explore these factors and further disentangle and explain the dynamics of human communication in social networks.

\section*{Methods}
\paragraph{Data.}
All data was collected in accordance with the Danish regulations for personal data; additionally the study has been subject to Institutional Review Board (IRB) approval. The IRB grantee is Indiana University (protocol number 1410501891), which was the hosting institution of the only U.S.-based author (Emilio Ferrara) at the time when this experiment was performed. All co-authors aligned to the requirements imposed by Indiana University’s approved protocol.
All data are available from the corresponding author on request.

\paragraph{Botnet creation.}
We designed the Twitter bots as part of a graduate course on social networks.
The goal was to create bots which appear, at a cursory glance, to be human-operated Twitter accounts, but in reality are algorithmically driven (by means of Python scripts).
The bot creation was divided into two phases: first, the goal was to build convincing accounts that real users might want to follow. Second, we worked to infiltrate a set of geographically co-located real users and spread new hashtags among them.

In phase 1, each group of 2-4 students manually created 1-3 personas (with interests, music taste, favorite sports team, etc.) and corresponding Twitter profiles, each with a profile picture, profile description, background picture, etc., resulting in a total of 39 bots.
Each group also manually posted a number of initial tweets for each bot.

One of the key objectives was to achieve a large follower base while maintaining a low following/follower ratio.
A low following/follower ratio is unusual among bots~\cite{subrahmanian2016darpa} and signals popularity on Twitter. 
Our bots achieved a low ratio by capitalizing on the fact that many new users with relatively few followers (and other Twitter bots) tend to reciprocate the link when they gain a new follower.  
Therefore, we used the following strategy:
Every day, each bot automatically followed approximately 100-200 randomly selected accounts with a low follower count or the string `followback' in the description.
After 24 hours, the bots unfollowed the accounts that failed to reciprocate their follow. 
This routine was repeated every subsequent day.
Using this strategy, the bots were able to maintain a following/follower ratio close to 1, while gaining large amounts of followers.
The bots avoided automatic detection by limiting the churn among their followers, since performing too many (un)follow operations in a day leads to a suspension of the account.
As a whole, the botnet was successful in gaining a large group of followers which grew steadily throughout the duration of the experiment, as shown in Figure~\ref{fig:followers_vs_time}A.

While attracting followers, the bots gradually assumed a number of behaviors designed to emulate human behavior:
\begin{description}
	\item \textit{Geographical patterns.} All bots' self-reported location in their Twitter profile was set to the San Francisco Bay Area. In addition, all bots tweeted with geo-tagged tweets, set to originate from a random location within the Bay Area bounding box. This allowed our bots to target a geographically-confined region.
	\item \textit{Temporal patterns.} Bots also timed their tweets to match typical diurnal patterns corresponding to the pacific time zone, and produce content that reflected circadian patterns of activity commonly observed online~\cite{golder2011diurnal}.
	\item \textit{Content.} Finally, based on simple natural language processing rules, the bots automated tweeting and re-tweeting of content that matched the persona developed above. 
\end{description}
As final step of phase 1, the bots unfollowed users which were obviously spam/bot accounts in order to decrease their following/followed ratio.
To investigate the quality of each bot, we routinely used the online service \textit{Bot or Not} API~\cite{davis2016botornot}\footnote{http://truthy.indiana.edu/botornot/} to ensure that the bots appeared human to state-of-the-art bot-detection-software.

In phase 2, the bots began following non-bot Twitter accounts within the target area (San Francisco/Bay Area), leveraging the information users self-reported in their Twitter profiles (location string).
To achieve the goal of having individuals in the target area following multiple bots, the bots maintained a shared list of Twitter accounts that followed-back any of the bots -- and all bots followed those real accounts over the following days.
As a result, many Twitter users in the target set ended up following multiple bots by the time when the interventions occurred during the period between November 15th to December 2nd, 2014.
The distribution of the number of bots followed by other Twitter users during the intervention period is shown in Figure~\ref{fig:followers_vs_time}B.

\paragraph{Statistics of observed data.}
The following shows how the observations, including the error bars, in figure \ref{fig:sc_cc_fit} were obtained.
For both SC and CC, we investigate how $P(\textrm{RT})$ changes as a function of $k$, then iterate over each of the interventions and for each target user we compute the distribution of exposure numbers, according to equation~\eqref{eqn:p_k_sc} for SC, and according to the Poisson binomial distribution shown in equation~\eqref{eqn:p_k_cc} for CC.
These distributions allow us to estimate the number of retweets after $k$ exposures in the following way: 
Consider a series of events $S_1, S_2, \ldots, S_n$, each representing a user retweeting an intervention-related tweet.
For an event $S_i$, we have probabilities $p_{i,1}, p_{i,2}, \ldots, p_{i,n}$ of the event representing $k=1, k=2, \ldots, k=n$ true exposures. 
Hence, considering a discrete value $k = j$, the event can belong to bin $j$ with a probability $p_{i,j}$, and it can belong in another bin with probability $1 - p_{i,j}$; i.e., it is drawn drawn from a Bernoulli distribution with $p_i = p_{i, j}$ and $\sigma_i^2 = p_{i,j}(1 - p_{i,j})$.
Similarly, the following event is drawn from another Bernoulli distribution independent of the first, and so the distribution of each bin follows another Poisson binomial distribution with $\mu = \sum_i p_i$ and $\sigma^2 = \sum_i \sigma_i^2$.
This process approaches the normal distribution $\mathcal{N}(\mu, \sigma^2)$, when the number of Bernoulli draws becomes large due to the central limit theorem (see SI Appendix for details). 
Thus, can we obtain an approximate distribution for the number of observed retweets for each value of $k$.

\paragraph{Bayesian Information Criterion.}
The Bayesian information criterion (BIC) score is defined as
\begin{equation}
\mathrm{BIC} = -2\ln(L) + k\ln(n),
\end{equation}
where $L$ is the likelihood of the data given the model, $k$ is the number of model parameters, and $n$ is the number of data points. We compute the likelihood based on the fits to the number of retweets, i.e. fits like those shown in Figure \ref{fig:sc_cc_fit}B: For each exposure number $k$, we have (from our previous analysis) an estimate of the number of times, $N_k$, a user has experienced $k$ exposures. To ensure a discrete number of retweets, we run a series of simulations, computing $P(k|A)$ for each retweeting user and adding $1$ to a bin $k$, which is selected using that probability distribution. We denote the number of retweets in bin $k$ by $n_k$, and discard bins in which $n_k < 5$. As our models provide the probability $P(\textrm{RT}|k)$ of each exposure succeeding in eliciting a response from the exposed user, the likelihood of each bin in one such simulation is given by a binomial distribution, and the total likelihood is simply the product of those, i.e.
\begin{equation}
L = \prod_{k} \binom{N_k}{n_k} P(\textrm{RT}|k)^{n_k}(1-P(\textrm{RT}|k))^{N_k-n_k}.
\end{equation}
We repeat this simulation $10^3$ times for both SC and CC for the full range of values of $q$.

\section*{Author contributions statement}
All authors contributed to writing the manuscript. SL, PS \& EF Designed the study, PS Collected the data, BM Developed statistical models and analyzed the data.

\begin{center}
	\Huge Evidence of Complex Contagion of Information in Social Media: An Experiment Using Twitter Bots\\
	\vspace*{-5mm}\mbox{}\\
	\huge \textit{Supplementary Information Appendix}\\
	\vspace*{2mm}\mbox{}\\
	\large Bjarke Mønsted, Piotr Sapieżyński, Emilio Ferrara \& Sune Lehmann
\end{center}
\pagebreak

\section*{Note on Generating Unique Unordered Subsets}
In the following, we outline an efficient method to compute the Poisson binomial distributions over $\kappa$ for the complex contagion model. The probability distributions
\begin{equation}
P(\kappa | A) = \sum_{s\in S_\kappa} \prod\limits_{i\in s}\left(1 - (1-q)^{a_i}\right)\prod\limits_{i \notin s} (1-q)^{a_i},
\end{equation}
We observe that while each element of an instance of
\begin{equation}
S_\kappa = \left\{s \subseteq \left\{1, 2, \ldots, |A| \right\}, |s| = \kappa \right\},
\end{equation}
the subsets of an activity list $A$ of size $k$, give a contribution to $P(\kappa | A)$, the contribution depends only on the elements of the set, and not on their order. As the number of these subsets is a combinatorial expression that grows very quickly with $|A|$ and $k$, and as the majority of the elements in $A$ were ones (as most of the bots only participated once in each intervention), we were able to complete the otherwise infeasible exact computation of $P(\kappa | A)$ by devising a method to generate unique unordered subsets from $A$ and then multiply each unique subset with its multiplicity, which could in turn be computed from simple combinatorial expressions.

To illustrate our approach, we first explain a common way (as implemented in the itertools module in Python 2.7) of generating all possible subsets of a given length of a set. As an example, we use $A = [2,1,2,1,3,1,1,4]$ and $k = 3$. Three pointers are then initialized to the three first values, and the last of those is set as the `active' pointer. A series of steps is then repeated until none are possible:
\begin{itemize}
	\item Attempt to move the active pointer one step to the right.
	\item If the pointer falls off the array or runs into another pointer, then set to active the pointer to the left of the current one and attempt again.
	\subitem$\circ$ Terminate if we run out of pointers, i.e. if no pointers can be moved anymore.
	\item When move is successful, generate the set of the values pointed to, move all pointers to the right of the active pointer to the positions immediately following it, and reset the `active' status to the rightmost pointer.
\end{itemize}
This then generates subsets like $(2, 1, 2)$, $(2, 1, 1)$, $\ldots$ $(2, 1, 4)$, $(2, 2, 1)$, $(2, 2, 3)$ etc. until $(1, 1, 4)$ where no more pointers are able to move and the procedure terminates.

This approach has the disadvantage of recounting subsets that are identical or are permutations of each other, such as $(2,1,2)$ and $(2,2,1)$ above. We remedy this by adding to the above algorithm a preprocessing step in which the input list $A$ is sorted, and then defining a new list $S$ in which the $i$'th element denotes the position the active pointer should be moved to given that it is pointing to $A[i]$ now, i.e. the value of $S[i]$ is the value of the first index $i'$ at which $A[i'] > A[i]$. Reusing the example above to illustrate this,
\begin{align*}
S &= [4, 4, 4, 4, 6, 6, 7, 8], \\
A &= [1, 1, 1, 1, 2, 2, 3, 4],
\end{align*}
where the final index $i = 8$ falls off the array, consistent with the description above. This generates subsets like $(1,1,1)$, $(1, 1, 2)$, $(1, 1, 3)$, $\ldots$, $(1, 2, 2)$, $(1,2,3)$ etc. until $(2,3,4)$. The multiplicity of each such subset can be computed analytically, allowing one to compute the probability of drawing each subset.

\begin{figure*}[t!]
	\centering
	\includegraphics[width = 0.7\textwidth]{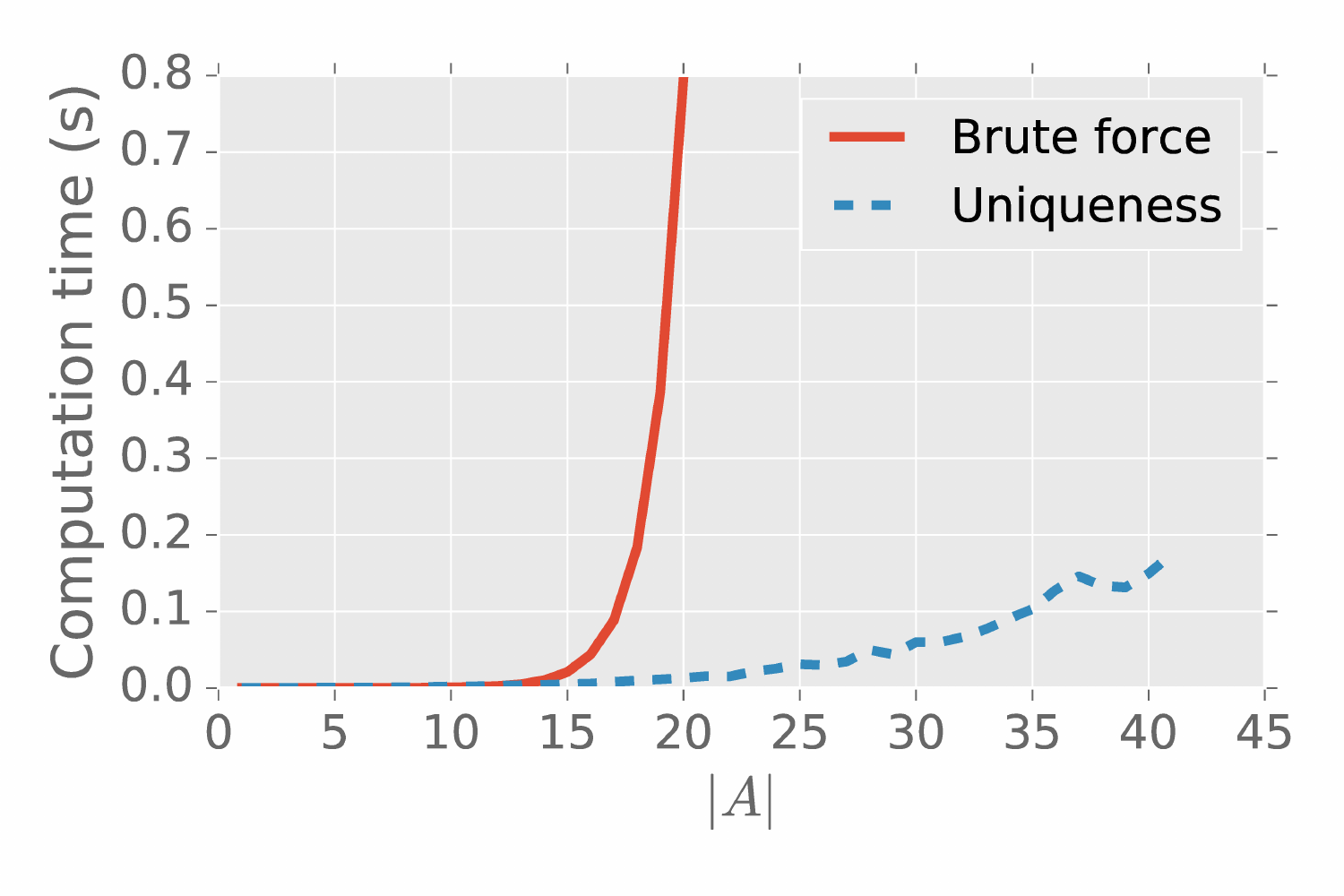}
	\caption{Time needed to compute $P(\kappa | A)$ using a brute force approach (solid red line) and our unique subset generation approach (dashed blue line). For each data point, we generated a simulated activity list with length $|A|$ with elements drawn with probabilities based on their frequency in our data, and computed $P(\kappa | A)$ using the two methods. The runtime is then averaged over 100 such computations in order to minimize noise. Our approach allows us to compute Poisson binomial distributions for activity lists of lengths that would otherwise be infeasible.}
	\label{fig:runtime}
\end{figure*}

When some elements of $A$ occur very frequently, this approach, which we call the `uniqueness' approach here as it only counts unique combinations, results in a clear improvement over the brute force approach. Indeed, for our data this approach turned out necessary to perform an otherwise infeasible computation. To illustrate this, we ran both method on simulated lists $A$ of varying lengths, constructing each list $A$ by randomly drawing elements from a distribution representative of our data. The resulting runtimes are shown in figure \ref{fig:runtime} and clearly show the speedup.

\section*{Note on the Gaussian approximation of the Poisson Binomial}
In the following, we detail why the Poisson binomial distribution consisting of $n$ draws fr	om separate Bernoulli distributions with individual probabilities $p_1, p_2, \ldots, p_n$, and accordingly, variances $\sigma_i^2 = p_i(1 - p_i)$ approaches $\mathcal{N}(\mu, \sigma^2)$, where $\mu = \sum_i p_i$ and $\sigma^2 = \sum_i p_i(1 - p_i)$ for large $n$. This closely follows the proof of the central limit theorem using characteristic functions, but we include the derivation in case some readers are unfamiliar with the proof.

The characteristic function (c.f.) of a random variable $X$ is defined as
\begin{equation}
\varphi_X(t) = \expect{e^{itX}} = \sum_{n=0}^\infty i^n \frac{\expect{X^n}}{n!}t^n, \label{eqn:charfunc}
\end{equation}
which results in the properties
\begin{align}
\varphi_{X + Y}(t) &= \varphi_X(t)\cdot \varphi_y(t), \label{eqn:cf_add} \\
\varphi_{cX}(t) &= \varphi_X(ct). \label{eqn:cf_mul}
\end{align}
Note that the c.f. for a Gaussian with zero mean, $\mathcal{N}(0, \sigma^2)$, is $\mathrm{e}^{-\sigma^2t^2/2}$. Rather than considering directly the random variables $X_i$ drawn from the Poisson binomial, we first subtract from each the mean success probability, i.e. we transform them like $X_i \rightarrow X_i - \overline{p}$. This subtraction ensures that $\expect{X_i} = 0$ and $\expect{X_i^2} = \sigma_i^2$, which simplifies the proof We also divide by the square root of the number of draws, as this makes the convergence easier to show. When the derivation is done, we can simply obtain the distribution of the sum by substituting $\mu \rightarrow \mu + \sum_i p_i$ and $\sigma \rightarrow \sqrt{n}\sigma$.

Writing out the c.f. for these random variables using \eqref{eqn:cf_add} and \eqref{eqn:cf_mul} gives
\begin{align}
\varphi_{\sum_i X_i/\sqrt{n}}(t) = \prod_{i = 1}^n \varphi_{X_i/\sqrt{n}}(t) = \prod_{i = 1}^n \varphi_{X_i}(t/\sqrt{n})
\end{align}
Using the properties following from the variables having zero mean, the expansion in \eqref{eqn:charfunc} becomes
\begin{equation}
\varphi_{X}(t) = 1 - \frac{\sigma^2}{2}t^2 + \mathcal{O}(t^3),
\end{equation}
so we get
\begin{equation}
\varphi_{\sum_i X_i/\sqrt{n}}(t) = \prod_{i = 1}^n \pp{1 - \frac{\sigma_i^2}{2n}t^2 + \mathcal{O}\pp{t^3 / \sqrt{n}^3}}.
\end{equation}
Using the expansion of the natural log, $\ln(1 - x) = \sum_{n= 1}^\infty\pp{-1}^{n}\frac{x^n}{n} = -x + \mathcal{O}(x^2)$, this may be rewritten as
\begin{align}
\varphi_{\sum_i X_i/\sqrt{n}}(t)
&= \mathrm{e}^{\ln\pp{\prod_{i = 1}^n \pp{1 - \frac{\sigma_i^2}{2n}t^2 + \mathcal{O}\pp{t^3 / \sqrt{n}^3}}}}, \\
&= \mathrm{e}^{\sum_{i = 1}^n \pp{-\frac{\sigma_i^2}{2n}t^2 + \mathcal{O}\pp{t^3 / \sqrt{n}^3}}}
\end{align}
Noting that $\sum_{i = 1}^n \sigma_i^2= n\cdot \overline{\sigma^2}$, meaning that when the moments of $X$ are bounded, the c.f. approaches
\begin{equation}
\lim_{n\rightarrow \infty} \varphi_{\sum_i X_i/\sqrt{n}}(t) = \mathrm{e}^{-\overline{\sigma^2}t^2/2},
\end{equation}
which is the c.f. for $\mathcal{N}(0, \overline{\sigma^2})$, meaning that the distribution goes to $\frac{1}{\sqrt{2\pi\overline{\sigma^2}}}\mathrm{e}^{-x^2/2\overline{\sigma^2}}$ as previously described, the distribution for the sum of the original random variables can then be obtained by substituting $x \rightarrow x - \sum_i p_i$ and $\overline{\sigma^2} \rightarrow n\cdot \overline{\sigma^2} = \sum_i p_i(1 - p_i)$.

\section*{Note on Data Collection}
The bots were hosted on Amazon EC2 instances, ensuring uninterrupted operation. Each team collected the data about their bot’s activity as well as the development of its Twitter network.  A script installed by each team facilitated the data collection effort. Once per hour it queried the Twitter API for the list of friends and followers of the team’s bot, as well as the content it recently (re)tweeted. The script would then send the data to the central server at DTU for storage and further analysis. By relegating the collection effort to each team we avoided the Twitter API call quotas, which would curb the process if we had used a single account to gather the data.

During the intervention period, another script, installed at the central server, used the Twitter Streaming API to collect tweets with predefined hashtags coming from all Twitter users. We combine the information collected by the teams with the record of all interventions-related tweets to analyze the dynamics of information spread.

\end{document}